\documentclass{aa}
\usepackage{graphics,epsf,epsfig,graphicx,natbib}
\bibpunct{(}{)}{;}{a}{}{,}

\begin{document}
\newcommand{\MHI}{\mbox{$M_{\mbox{\tiny HI}}$}}
\newcommand{\NHI}{\mbox{$N_{\mbox{\tiny HI}}$}}
\newcommand{\VHI}{\mbox{$V_{\mbox{\tiny HI}}$}}
\newcommand{\Vopt}{\mbox{$V_{\mbox{\tiny opt}}$}}
\newcommand{\NH}{\mbox{$N_{\mbox{\tiny H}}$}}
\newcommand{\Lb}{\mbox{$L_{\mbox{\tiny B}}$}}
\newcommand{\LbA}{\mbox{$L_{\mbox{\tiny B}}^{\mbox{\tiny A}}$}}
\newcommand{\LbB}{\mbox{$L_{\mbox{\tiny B}}^{\mbox{\tiny B}}$}}
\newcommand{\Lv}{\mbox{$L_{\mbox{\tiny V}}$}}
\newcommand{\Lr}{\mbox{$L_{\mbox{\tiny R}}$}}
\newcommand{\LHa}{\mbox{$L_{\mbox{\tiny H}\alpha}$}}
\newcommand{\Ha}{\mbox{H$_{\alpha}$}}
\newcommand{\Hb}{\mbox{H$_{\beta}$}}
\newcommand{\Hg}{\mbox{H$_{\gamma}$}}
\newcommand{\Hd}{\mbox{H$_{\delta}$}}
\newcommand{\Mb}{\mbox{$M_{\mbox{\tiny B}}$}}
\newcommand{\Mv}{\mbox{$M_{\mbox{\tiny V}}$}}
\newcommand{\Mr}{\mbox{$M_{\mbox{\tiny R}}$}}
\newcommand{\MHH}{\mbox{$M_{\mbox{\tiny H$_2$}}$}}
\newcommand{\HH}{\mbox{H$_2$}}
\newcommand{\Mo}{\mbox{M$_{\odot}$}}
\newcommand{\Mvir}{\mbox{$M_{\mbox{\tiny vir}}$}}
\newcommand{\Mdyn}{\mbox{$M_{\mbox{\tiny dyn}}$}}
\newcommand{\Mmar}{\mbox{$M_{\mbox{\tiny mar}}$}}
\newcommand{\Lo}{\mbox{L$_{\odot}$}}
\newcommand{\Zo}{\mbox{Z$_{\odot}$}}
\newcommand{\x}{\mbox{$\times$}}
\newcommand{\Dph}{\mbox{$D_{25}$}}
\newcommand{\Vrot}{\mbox{$V_{\mbox{\tiny rot}}$}}
\newcommand{\rH}{\mbox{$r_{\mbox{\tiny H}}$}}
\newcommand{\Mk}{\mbox{$M_{\mbox{\tiny K}}$}}
\newcommand{\COa}{\mbox{CO(1$\rightarrow$0)}}
\newcommand{\COb}{\mbox{CO(1$\rightarrow$1)}}
\newcommand{\CO}{\mbox{CO}}
\newcommand{\cmm}{\mbox{cm$^{-2}$}}
\newcommand{\cmmm}{\mbox{cm$^{-3}$}}
\newcommand{\Mpc}{\mbox{Mpc}}
\newcommand{\kms}{\mbox{km~s$^{-1}$}}
\newcommand{\pcc}{\mbox{pc$^{-2}$}}
\newcommand{\kpc}{\mbox{kpc}}
\newcommand{\sbr}{\mbox{mag/\fbox{}\arcsec}}
\newcommand{\sbB}{\mbox{$\mu _{\mbox{\tiny B}}$}}
\newcommand{\sbV}{\mbox{$\mu _{\mbox{\tiny V}}$}}
\newcommand{\sbc}{\mbox{$\mu _{\mbox{\tiny B}0}$}}
\newcommand{\fB}{\mbox{$f_{\mbox{\tiny B}}$}}
\newcommand{\aop}{\mbox{$a_{25}$}}
\newcommand{\bop}{\mbox{$b_{25}$}}
\newcommand{\tabsp}{\noalign{\smallskip}}
\newcommand{\BV}{\mbox{$\mbox{B} - \mbox{V}$}}
\newcommand{\VK}{\mbox{$\mbox{V} - \mbox{K}$}}
\newcommand{\sigint}{\mbox{$\sigma _{\mbox{\tiny int}}$}}
\newcommand{\Hoa}{\mbox{H$_0= 75$~km~s$^{-1}$~Mpc$^{-1}$}}
\newcommand{\Hob}{\mbox{H$_0= 70$~km~s$^{-1}$~Mpc$^{-1}$}}
\newcommand{\qo}{\mbox{q$_0$}}
\newcommand{\Mbo}{\mbox{$M_{\odot}^{\mbox{\tiny B}}$}}
\newcommand{\Vpar}{\mbox{$V_{\mbox{\tiny par}}$}}
\newcommand{\eqw}{\mbox{$W$(H$_{\beta}$)}}
\newcommand{\Ab}{\mbox{$A_{\mbox{\tiny B}}$}}
\newcommand{\Av}{\mbox{A$_{\mbox{\tiny V}}$}}
\newcommand{\noteb}{\mbox{$^{\mbox{\tiny +}}$}}
\newcommand{\mJyb}{\mbox{mJy~beam$^{-1}$}}
\newcommand{\mJy}{\mbox{mJy}}
\newcommand{\K}{\mbox{K}}
\newcommand{\Ts}{\mbox{$T_{\mbox{s}}$}}
\newcommand{\TOIII}{\mbox{$T_e({\mbox{OIII}}$)}}
\newcommand{\TOII}{\mbox{$T_e({\mbox{OIII}}$)}}
\newcommand{\micron}{\mbox{$\mu$m}}
\newcommand{\Oabun}{\mbox{$12 + \log(\frac{\mbox{O}}{\mbox{H}})$}}
\newcommand{\er}{\mbox{$\pm$}}

\newcommand{\fb}{\mbox{$f_{\mbox{\tiny B}}$}}
\newcommand{\Lfir}{\mbox{$L_{\mbox{\tiny FIR}}$}}
\newcommand{\Lir}{\mbox{$L_{\mbox{\tiny IR}}$}}
\newcommand{\uflux}{\mbox{erg~cm$^{-2}$~s$^{-1}$}}
\newcommand{\usbflux}{\mbox{erg~s$^{-1}$~cm$^{-2}$~arcsec$^{-2}$}}
\newcommand{\ufluxm}{\mbox{erg~cm$^{-2}$~s$^{-1}$~\AA$^{-1}$}}
\newcommand{\ul}{\mbox{erg~s$^{-1}$}}
\newcommand{\uSFR}{\mbox{M$_{\odot}$~yr$^{-1}$}}
\newcommand{\muJy}{\mbox{$\mu$Jy}}
\newcommand{\OIIIa}{\mbox{[OIII]$_{\lambda 4959}$}}
\newcommand{\OIIIb}{\mbox{[OIII]$_{\lambda 5007}$}}
\newcommand{\OIIIc}{\mbox{[OIII]$_{\lambda 4363}$}}
\newcommand{\OI}{\mbox{[OI]$_{\lambda 6300}$}}
\newcommand{\OII}{\mbox{[OII]$_{\lambda 3727}$}}
\newcommand{\OIIb}{\mbox{[OII]$_{\lambda 7320,30}$}}

\newcommand{\OIIIt}{\mbox{[OIII]}}
\newcommand{\OIIt}{\mbox{[OII]}}
\newcommand{\OIt}{\mbox{[OI]}}
\newcommand{\SIIt}{\mbox{[SII]}}
\newcommand{\NIIt}{\mbox{[NII]}}
\newcommand{\NIt}{\mbox{[NI]}}
\newcommand{\ArIIIt}{\mbox{[ArIII]}}

\newcommand{\NIIa}{\mbox{[NII]$_{\lambda 6548}$}}
\newcommand{\NIIb}{\mbox{[NII]$_{\lambda 6584}$}}
\newcommand{\SIIa}{\mbox{[SII]$_{\lambda 6717}$}}
\newcommand{\SIIb}{\mbox{[SII]$_{\lambda 6731}$}}
\newcommand{\SII}{\mbox{[SII]$_{\lambda 6717,6731}$}}

\newcommand\cola {\null}
\newcommand\colb {&}
\newcommand\colc {&}
\newcommand\cold {&}
\newcommand\cole {&}
\newcommand\colf {&}
\newcommand\colg {&}
\newcommand\colh {&}
\newcommand\coli {&}
\newcommand\colj {&}
\newcommand\colk {&}
\newcommand\coll {&}
\newcommand\colm {&}
\newcommand\coln {&}
\newcommand\eol{\\}
\newcommand\extline{&&&&&&&&&\eol}

%
\def\HI{H\,{\small I}}
\def\HI{H\,{\sc i}}
\def\HII{H\,{\sc ii}}
\def\HIit{\mbox{H\hspace{0.155 em}{\footnotesize \it I}}}
\def\nan{Nan\c{c}ay}
\def\sbu{${\rm mag\,\,arcsec^{-2 }} $ \ }
\newcommand{\am}[2]{$#1'\,\hspace{-1.7mm}.\hspace{.0mm}#2$}
\newcommand{\as}[2]{$#1''\,\hspace{-1.7mm}.\hspace{.0mm}#2$}
\newcommand\btab[5]{\begin{table*}[#1]{\parbox{#4}{\caption{#2}}\rule[-0.5ex]{0cm}{0.5ex} }
\begin{tabular*}{#4}{#5} }
\newcommand\sbtab[5]{\begin{table}[#1]{\parbox{#4}{\caption{#2}}\rule[-0.5ex]{0cm}{0.5ex} }
\begin{footnotesize}
\begin{tabular*}{#4}{#5} }
\newcommand{\etab}[4]{
\end{tabular*}
\vspace*{#1}
\begin{flushleft}
\parbox{#2}{#3}
\end{flushleft}
\label{#4}
\end{table*} }
\def\kato{\rule[-1.25ex]{0cm}{1.25ex}}
\def\pano{\rule[0.0ex]{0cm}{2.5ex}}
%
%
%
%
%
%


\def\rf@jnl#1{{#1}}
\def\aj{\rf@jnl{AJ }}                   
\def\araa{\rf@jnl{ARA\&A }}             
\def\apj{\rf@jnl{ApJ }}                 
\def\apjl{\rf@jnl{ApJ }}                
\def\apjs{\rf@jnl{ApJS }}               
\def\ao{\rf@jnl{Appl.~Opt.}}           
\def\apss{\rf@jnl{Ap\&SS }}             
\def\aap{\rf@jnl{A\&A }}                
\def\aapr{\rf@jnl{A\&A~Rev.}}          
\def\aaps{\rf@jnl{A\&AS }}              
\def\azh{\rf@jnl{AZh }}                 
\def\baas{\rf@jnl{BAAS }}               
\def\jrasc{\rf@jnl{JRASC }}             
\def\memras{\rf@jnl{MmRAS }}            
\def\mnras{\rf@jnl{MNRAS }}             
\def\pra{\rf@jnl{Phys.~Rev.~A}}        
\def\prb{\rf@jnl{Phys.~Rev.~B}}        
\def\prc{\rf@jnl{Phys.~Rev.~C}}        
\def\prd{\rf@jnl{Phys.~Rev.~D}}        
\def\pre{\rf@jnl{Phys.~Rev.~E}}        
\def\prl{\rf@jnl{Phys.~Rev.~Lett.}}    
\def\pasp{\rf@jnl{PASP }}               
\def\pasj{\rf@jnl{PASJ }}               
\def\qjras{\rf@jnl{QJRAS }}             
\def\skytel{\rf@jnl{S\&T }}             
\def\solphys{\rf@jnl{Sol.~Phys.}}      
\def\sovast{\rf@jnl{Soviet~Ast.}}      
\def\ssr{\rf@jnl{Space~Sci.~Rev.}}     
\def\zap{\rf@jnl{ZAp }}                 
\def\nat{\rf@jnl{Nature }}              
\def\iaucirc{\rf@jnl{IAU~Circ.}}       
\def\aplett{\rf@jnl{Astrophys.~Lett.}} 
\def\apspr{\rf@jnl{Astrophys.~Space~Phys.~Res.}}
\def\bain{\rf@jnl{Bull.~Astron.~Inst.~Netherlands}} 
\def\fcp{\rf@jnl{Fund.~Cosmic~Phys.}}  
\def\gca{\rf@jnl{Geochim.~Cosmochim.~Acta}}   
\def\grl{\rf@jnl{Geophys.~Res.~Lett.}} 
\def\jcp{\rf@jnl{J.~Chem.~Phys.}}      
\def\jgr{\rf@jnl{J.~Geophys.~Res.}}    
\def\jqsrt{\rf@jnl{J.~Quant.~Spec.~Radiat.~Transf.}}
\def\memsai{\rf@jnl{Mem.~Soc.~Astron.~Italiana}}
\def\nphysa{\rf@jnl{Nucl.~Phys.~A}}   
\def\physrep{\rf@jnl{Phys.~Rep.}}   
\def\physscr{\rf@jnl{Phys.~Scr}}   
\def\planss{\rf@jnl{Planet.~Space~Sci.}}   
\def\procspie{\rf@jnl{Proc.~SPIE}}   

\let\astap=\aap
\let\apjlett=\apjl
\let\apjsupp=\apjs
\let\applopt=\ao

\title{A top-down scenario for the formation of massive Tidal Dwarf Galaxies}
        
\author{Pierre-Alain Duc \inst{1,2}, Fr\'ed\'eric Bournaud \inst{3,1,4} \& Fr\'ed\'eric Masset \inst{1}}
 
\offprints{P.--A. Duc, \email{paduc@cea.fr}} 
\institute{ CEA/DSM/DAPNIA, Service d'Astrophysique, Saclay, 91191 Gif sur Yvette Cedex, France
\and
CNRS, FRE 2591
\and 
Observatoire de Paris, LERMA, 61 Av. de l'Observatoire, 75014 Paris, France
\and
Ecole Normale Sup\'erieure, 45 rue d'Ulm, 75005 Paris, France
}

\date{Accepted for publication in A\&A}
\authorrunning{P.--A. Duc et al.} 
\titlerunning{Formation of massive Tidal Dwarf Galaxies}

\abstract{ 
Among those objects formed out of collisional debris during galaxy mergers, the prominent
gaseous accumulations observed near the tip of some long tidal tails are the
most likely to survive long enough to form genuine recycled galaxies.
Using simple numerical models, Bournaud, Duc \& Masset (2003) claimed that 
tidal objects as massive as  $10^9~\Mo$ could only form, in these simulations, within 
extended dark matter (DM) haloes. We present here a new set of  simulations 
of galaxy collisions to further investigate the structure of tidal tails. First of all,
we checked that massive objects are still produced in full N-body codes that
include feedback  and a large number of particles. Using a simpler N-body code with 
rigid haloes, we 
noticed that  dissipation and self-gravity in the tails, although important, are not the key 
factors. Exploiting toy models, we found that, for truncated DM haloes, 
material is stretched along the tail, while, within extended haloes,  the tidal field 
can efficiently carry away from the disk a large fraction of the gas, while maintaining
 its surface density to a high value. This creates a density enhancement near the tip
of the tail. Only later-on, self-gravity takes over; the gas clouds collapse and start
forming stars. Thus, such  objects  were
fundamentally formed following a kinematical process, according to a top-down 
scenario, contrary to the less massive Super Star Clusters that are also
present around mergers. This conclusion leads us to introduce a restrictive 
definition for Tidal Dwarf Galaxies (TDGs) and their progenitors, considering only 
the most massive ones, initially mostly made of gas, that were able to pile up in 
the tidal tails. More simulations will be necessary to precisely determine the
fate of these proto--TDGs and estimate their number.
}

\maketitle


\section{Introduction}

Numerous papers have reported the formation of presumably bound stellar 
structures around interacting systems. Their blue colors and high metallicity indicate
that they were formed rather recently out of material processed in the disk of their parent galaxies.
The census of young objects in mergers is still on going and their taxonomy not yet completed. \\
- {\it Super Star Clusters (SSCs)}, with typical masses of $10^5~\Mo$, appear on optical images 
as compact stellar condensations. They are either located in the central regions of
mergers \citep[e.g.,][]{Holtzman92,Whitmore99,Zepf99}, in tidal bridges or
 along extended tidal tails 
(e.g., Gallagher et al. 2001; Knierman et al. 2003; Saviane, Hibbard \& Rich 2004)
\nocite{Gallagher01,Knierman03, Saviane04} and may be
dust enshrouded \citep{Gilbert00}.
 {\it Young Massive Clusters (YMCs)} were identified in high-resolution HST images
\citep[e.g.,][]{deGrijs03,Tran03}.
Their blue luminosity, inferred mass and size are such that some of them 
may evolve into Globular Clusters with typical masses of $10^6~\Mo$ \citep[e.g.,][]{Schweizer96}. 
Some $10^7~\Mo$ HI fragments discovered around mergers 
could be their gaseous progenitors \citep{English03}. \\
- {\it Giant HII Complexes (GHCs)}, i.e. star-forming regions with luminosities
exceeding the Giant HII Regions found in isolated spiral disks, were discovered in
 long optical tidal tails (e.g., Weilbacher, Duc \& Fritze-v. Alvensleben 2003; 
L{\' o}pez-S{\' a}nchez, Esteban \& Rodr{\'{\i}}guez 2004).
\nocite{Weilbacher03,Lopez04} Further away, {\it Intergalactic Emission Line
 Regions}, characterized by a very low underlying old stellar content, and  sometimes by their compact
aspect (the so-called {\it EL-Dots}) were detected in several clusters of galaxies. 
Their optical spectra are typical of star-forming HII regions and their rather high oxygen
abundances indicate that they are formed of pre-enriched gas probably stripped from
parent colliding galaxies \citep[e.g.,][]{Gerhard02,Cortese04,Ryan-Weber04,Mendes04}. \\
- Finally, the formation in tidal debris of {\it Tidal Dwarf Galaxies (TDGs)}, i.e. objects with apparent 
masses and sizes of dwarf galaxies, has been  reported in several studies of interacting systems
\citep[e.g.,][and references thereein]{Duc99b,Hibbard04}. 
They contain large quantities of gas in atomic, molecular and ionized form and
have luminous masses of typically $10^9~\Mo$ \citep[e.g.,][]{Braine01}. Because they are most 
often observed at or near the tip of long optical tidal tails, the very existence of such massive 
objects has been challenged. Indeed an apparent accumulation of tidal material could be artificial. 
 In 3--D space, tidal tails are curved. Seen edge-on, they appear as linear structures and may show 
at their tip fake mass concentrations caused by material projected along the line of sight \citep{Hibbard04,Mihos04}.
 Kinematical studies of tidal tails help in identifying such projection effects \citep{Bournaud04}. The 
spatial and velocity coincidence of the different phases of the tidal gas towards a TDG 
candidate adds circumstantial evidence as to its reality  \citep{Braine01}.\\

These objects around mergers all contain stars with  an age,  derived from their
colors or spectroscopic properties,  which is comparable to or 
smaller than the dynamical age 
of the collision between the parent galaxies. They were hence most probably formed during 
the merging process. This does not mean however that a single mechanism is at their origin.
Indeed,  the heterogeneity of their properties (gaseous content, stellar population, location,
and above all their luminosity/mass) rather suggests that several physical processes 
take place during the collision. The initial conditions of the encounter may also matter
to account, for instance, for the formation of the most massive tidal objects, exceeding 
$10^9~\Mo$, i.e. those often referred as Tidal Dwarf Galaxies.

Numerical simulations of collisions are particularly helpful to identify the mechanism 
responsible for the formation of young objects in tidal debris. Soon after the study by 
Mirabel, Dottori \& Lutz (1992) \nocite{Mirabel92} of a Tidal Dwarf Galaxy candidate in the Antennae system, \cite{Barnes92} 
and Elmegreen, Kaufman \& Thomasson (1993) \nocite{Elmegreen93}  published  numerical models exhibiting bound condensations 
along tidal tails. While in the  \cite{Barnes92} simulations, the condensations formed
from gravitational instabilities in the stellar component, those of  \cite{Elmegreen93}
were produced directly from gas clouds, the velocity dispersion of which had
increased due to the collision. Further simulations, made with a higher resolution and/or 
more  realistic initial conditions, especially regarding the respective distribution
of the gaseous and stellar components,   could produce along the tails objects likely
to be the progenitors of   SSCs or globular clusters \citep[e.g.,][]{Barnes96,Kroupa98,Bekki02}.
Their masses of typically $10^8~\Mo$ are however one order of magnitude 
lower than that measured in the Tidal Dwarf Galaxies located at the tip of tidal tails.

In a few older simulations, however, the formation of  massive accumulations of tidal material
were noted (See Fig.~3 in \cite{Barnes92c} or Fig.3b in  \cite{Elmegreen93}) but
were not much commented as they were quickly disrupted or went out of the
volume studied in the model.
 \cite{Elmegreen93} actually concluded that massive galaxy interactions
could lead to the ejection of the most outer HI disk, forming an ``extended $10^9~\Mo$ pool
at the end of the tidal tail''.
Recently, Bournaud, Duc \& Masset (2003) \nocite{Bournaud03} (hereafter BDM03) using a new set of simulations
involving simple co-planar collisions showed that such large end-of-tail blobs could 
condense and become gravitationally bound.
Varying the initial size of the dark matter (DM)
haloes in the parent galaxies, they found that objects as massive as $10^9~\Mo$, i.e. 
similar to TDGs, could only be produced when the haloes were extended enough -
at least ten times the optical radius. Most often, dark matter haloes are truncated in 
numerical simulations although cosmological models predict their large size. 
This trick saves a great number of DM particles and is justified as long as
the central regions of colliding galaxies are studied. BDM03 actually  found that 
the  extent of the DM halo influences the {\it internal structure of the tidal tails,
 i.e. the distribution of matter along them}, 
while previous  works had more focussed on the length and curvature
of these tails (e.g., Mihos, Dubinski \& Hernquist 1998; Springel \& White 1999; 
Dubinski, Mihos \& Hernquist 1999).
\nocite{Mihos98b,Springel99,Dubinski99}

\begin{table*}
\begin{minipage}{13cm}
\begin{tabular}{cccccccccc}
\hline
\hline
Run & $N$\footnote{Spatial resolution: $N$ is the number of cells  on the Cartesian grid. See text for the associated number of particles and softening length.} & $M$\footnote{Stellar mass ratio.} & $m_\mathrm{g}$\footnote{Gas mass fraction in the most massive galaxy; $m_g$ is the gas to disk+bulge mass ratio.} & $m_\mathrm{g}$\footnote{Gas mass fraction in the less massive galaxy} & $V$\footnote{Relative velocity of the two galaxies computed at an infinite distance in \kms.} & $b$\footnote{Impact parameter in kpc.} & $i_1$\footnote{Inclination of the most massive galactic disk with respect to the orbital plane.} & $i_2$\footnote{Inclination of the less massive galactic disk with respect to the orbital plane.} & $\omega$\footnote{Angle between the two galactic disks.} \\
\hline
A & $1024^3$ & 1.2 & 8\% & 10\%& 150 & 90 & 15 & 20 & 27 \\
B & $512^3$ & 1.7 & 5\% & 12\%& 150 & 50 & 45 & 20 & 40 \\
C & $512^3$ & 1.0 & 8\% & 8\% & 170 & 75 & 15 & 15 & 30 \\
D & $512^3$ & 1.3 & 9\% & 9\% & 120 & 120 & 25 & 25 & 35 \\
\hline
\end{tabular}
\end{minipage}
\caption{Run parameters}\label{param}
\end{table*}

The simulations presented by BDM03 had a number of limits. The DM and stellar
 components were computed with a 3--D FFT code at a resolution
of only 5 kpc while the gas was simulated in 2--D (i.e. in a co-planar encounter) with a
 resolution of 150 pc high enough to resolve the gaseous tidal tails. Thus, the possible role of
stellar sub-structures (which could not form in the low-resolution code) in building TDGs was
ignored. Beside, the gas/DM halo interaction was not taken into
account. In this paper, we check the results of BDM03 in
 full 3--D N-body simulations that have, beside, a much higher 
resolution in both the DM, stellar and gas components (Sect.~\ref{sect.:hires-sim}). 
Having confirmed with this new code the early results of BDM03 and assessed the fundamental role 
of kinematical processes in structuring tidal tails, we use simpler simulations and 
qualitative arguments to
clarify the  effect of tidal forces within extended DM haloes (Sect.~\ref{sect.:ingredients}).
Finally, in Sect.~\ref{sect.:discussion}, we revisit the definition of Tidal Dwarf Galaxies,
arguing that, being formed according to a top-down scenario, they have a different physical
origin than the other tidal objects.

This paper only discusses the physics of the formation of  massive TDGs. It does not address
their abundance nor their  evolution and survival time. Answering these issues  of
cosmological importance  requires to carry out a large number of simulations with a
comprehensive set of initial conditions. This work is in progress and will be presented elsewhere.


\section{Formation of TDGs in full 3--D N-body simulations}
\label{sect.:hires-sim}

We have carried out a series of full 3--D N-body simulations of the merger of two
spiral galaxies with mass ratios between  1:1 and 1.7:1  and different orbital parameters. 
Their resolution, which was much higher than in BDM03, allowed to precisely study 
the internal structure of the tidal tails.

\subsection{Code description}

\subsubsection{Physical model}
We have used the N-body FFT code of \cite{Bournaud03a}. In the main simulation presented in detail hereafter, the 
number of particles for the most massive galaxy was $2 \times 10^6$ for dark matter, $10^6$ for stars, and $10^6$ 
for gas. For the other galaxy, the numbers were scaled to its mass. The gravitational potential was computed on a
 Cartesian grid of size $1024^3$, with a  softening length of 390 pc. We have also run three other simulations with a four times smaller number of particles meshed on a $512^3$ grid, and a softening length of 780 pc. 

The dissipative nature of the ISM has been modeled through the sticky-particles code of \cite{Bournaud02},
 with elasticity parameters $\beta_t = \beta_r = 0.8$ for cloud-cloud collisions. Star formation was described by a 
generalized Schmidt law \citep{Schmidt59}, and we employed a time-dependent stellar mass-loss model inspired 
of that of Jungwiert, Combes \& Palou{\v s} (2001) \nocite{Jungwiert01} and described in \cite{Bournaud02}. We then computed the star formation and mass-loss 
rates in each cell of the Cartesian grid, according to the values of these two rates, converting a fraction 
of gaseous particles into stellar ones and vice versa.

\subsubsection{Mass distribution of galaxies}
Each galaxy was made-up of a spherical dark halo, a stellar and gaseous disk, and a spherical bulge. For the most 
massive galaxy, with a visible mass of $2\times 10^{11}$ M$_{\sun}$, the stellar disk was a Toomre disk with a radius of
 15 kpc, a radial scale-length of 5 kpc and  the same vertical distribution as in \cite{Bournaud02}. The gaseous disk is a Toomre disk of truncation radius 37.5 kpc and a radial scale-length of 15 kpc. Such a gaseous disk which is 2.5 times more extended than the stellar one is representative of most spiral galaxies \citep[e.g.,][]{Roberts94}. The bulge was modeled by a Plummer sphere of scale-length 1 kpc and truncation radius 3 kpc. The dark halo was a softened isothermal sphere truncated at radius 150 kpc (10 times the stellar disk radius). Its density profile is given by: 
\begin{equation}
\rho(r)=\frac{\sigma ^2}{2 \pi G (r^2 + r_c^2)}
\end{equation}
We used a core radius $r_c=4.5$ kpc. For the other galaxy, the radial distribution of matter was scaled by the square 
 root of its stellar mass. In all simulations, we chose a bulge to disk mass ratio of 0.3, and a dark-to-visible mass 
ratio inside the stellar radius of 0.6. The mass fraction of gas inside the disk has been varied
(see Table~\ref{param}).

\subsubsection{Galactic encounter and orbital parameters}
Table~\ref{param} lists the values of the orbital parameters used in our simulations.
The ratio $M$ of the stellar masses between the two galaxies roughly corresponds to the ratio of the
dark halo masses and hence of the total masses. We chose orbital parameters that favor the formation of a long tidal tail for at least 
one of the two galaxies; in particular we have only considered prograde  orbits.

\begin{figure*}
\centering
\centerline{\psfig{file=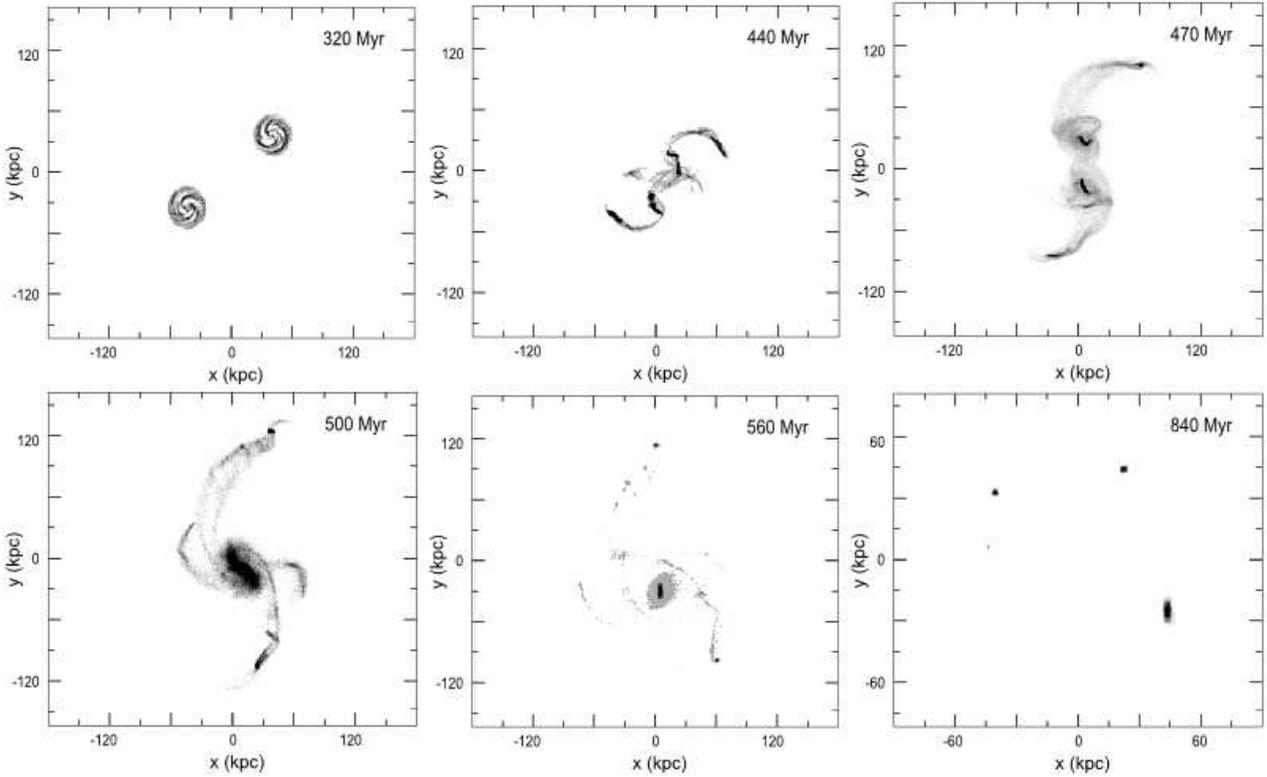,angle=-90,width=17cm}}
\caption{Evolution of the gas distribution for the full N--body simulations of run A. The system is seen face--on.
The gas is displayed with a logarithm intensity scale.}
\label{fig:simA} 
\end{figure*}

\subsection{Results}
\begin{figure}
\centering
\resizebox{8cm}{!}{\includegraphics{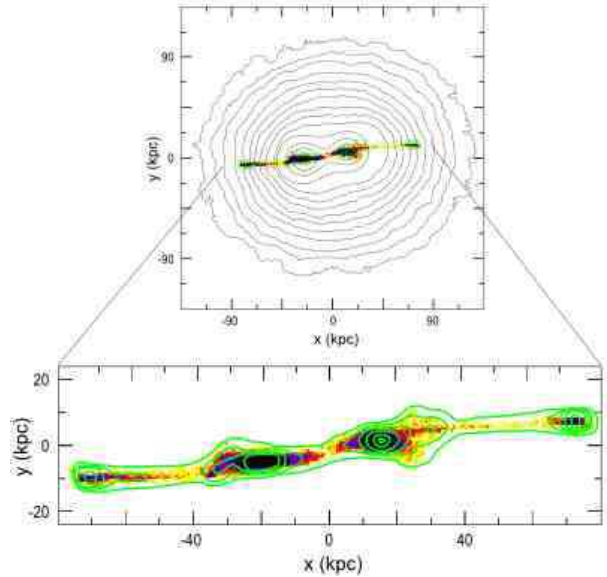}}
\caption{Bottom: contours of the gaseous component superimposed on a grayscale representation 
of the stellar component at $t=$ 470 Myr (simulation A). Top: zoom-out with the contours of the dark matter
haloes superimposed. The system is seen edge-on.}
\label{fig:simA:edge-on} 
\end{figure}

\begin{figure}
\centering
\resizebox{8cm}{!}{\includegraphics{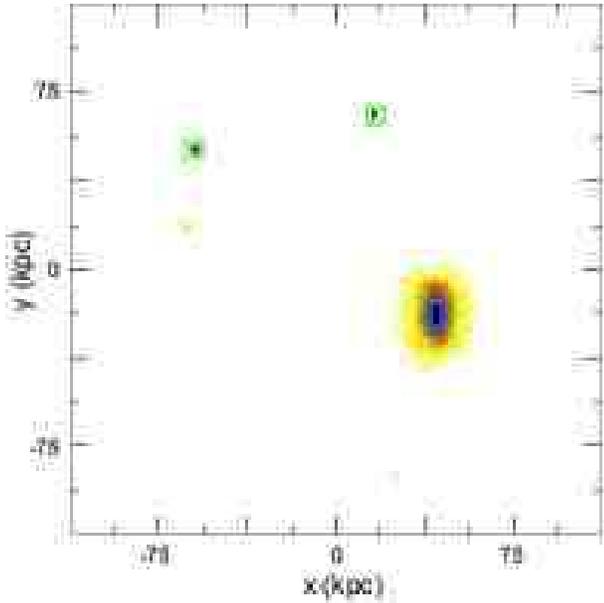}}
\caption{Contours of the gaseous component superimposed on a grayscale representation 
of the stellar component at $t=$ 850 Myr (run A). The system is seen face-on.}
\label{fig:simA:face-on} 
\end{figure}

Fig.~\ref{fig:simA} presents the evolution of the gas distribution for simulation A. 
The stellar, gaseous and dark matter components are shown at key moments of
the merging sequence in Fig.~\ref{fig:simA:edge-on} and Fig.~\ref{fig:simA:face-on}.

In run A, two major tidal tails of $\simeq$ 100 kpc developed soon after the periapse (at $t=$410 Myr).
At $t=$470 Myr, massive  gas clouds have piled up at the tip of each tidal
tail: they are the progenitors of the TDGs. We find that 70\% of the gas along the tails is located in the 20 last kpc, while the tails are about 100 kpc long \footnote{We computed the gas mass in a tail as the mass at radii larger than 30 kpc from the galaxy center. We chose to consider for the mass of the TDG progenitor all the matter included within a 
fixed aperture, rather than using a friend-to-friend algorithm that  would better delineate the bound tidal object. Indeed this 
is how the luminosity and mass are usually measured in real observations. Thus, the real mass  may have slightly
 been overestimated, as in the reality. The initial limit of 20 kpc was determined varying the aperture radius,
which is also the method used for aperture photometry.}. 
The total masses of the two TDG progenitors at $t=$470 Myr, computed in spheres of diameter 20 kpc centered on the TDG luminosity peak, are resp. $2.9 \x 10^9~\Mo$ and $2.3 \x 10^9~\Mo$ (see Table~\ref{tab:TDGs}). At this stage, most of the mass is in the gas; the stars contribute to 25\% of the mass. The majority of the stars were actually formed in situ in the tail. The old stellar population
 which was tidally ejected along with the gas  contributes to only resp. 5\% and 12\% of the total mass.

 At $t=$500 Myr, the stellar disks have already merged. The two major condensations at the tip of the tails have become more compact due to the action of self-gravity. Smaller sub-structures developed along the two major tails as well as in the two smaller counter-tails. They do not survive later on either because they are tidally disrupted or because they simply fall back. At $t=$840 Myr
 (430~Myr after the periapse), the two TDGs, which now appear isolated, orbit around the merger remnant at a radius of about 100 kpc. Their respective masses are now $2.2 \x 10^9~\Mo$ and $1.8 \x 10^9~\Mo$ \footnote{One fourth on the mass of the condensations of gas measured at $t=$470 Myr thus turned out  not to belong to the gravitationally linked object.}. At the end of the simulations, the TDGs have converted a large fraction of their gas into young stars (see Fig.~\ref{fig:simA:face-on}). The mass of old stars generally represents 5 to 15 \% of the mass of the  TDG (see Table~\ref{tab:TDGs}), but this quantity may vary with the extent of the gaseous disk compared to the stellar one, which has not been changed here. 

As many real interacting systems for which detailed studies of TDGs exist are 
observed close to edge-on, we present in Fig.~\ref{fig:simA:edge-on} an edge-on
view at a time when the TDGs have already formed. With this line of sight,
the tidal tails appear as linear structures and the TDGs as prominent 
overdensities at their apparent tip. This is close to what is observed in 
systems like NGC 4676 \citep{Hibbard96} or Arp 105 \citep{Duc97b}. The simulated systems seem
however to have more narrow tidal features, especially in the stellar 
component. The fact that the mass is dominated by the gaseous component
is also quite realistic. The HI mass of typical TDG candidates is indeed 
a factor of 2--5 higher than the stellar mass estimated from their optical luminosity \citep{Braine01}.

\begin{figure*}
\centerline{\psfig{file=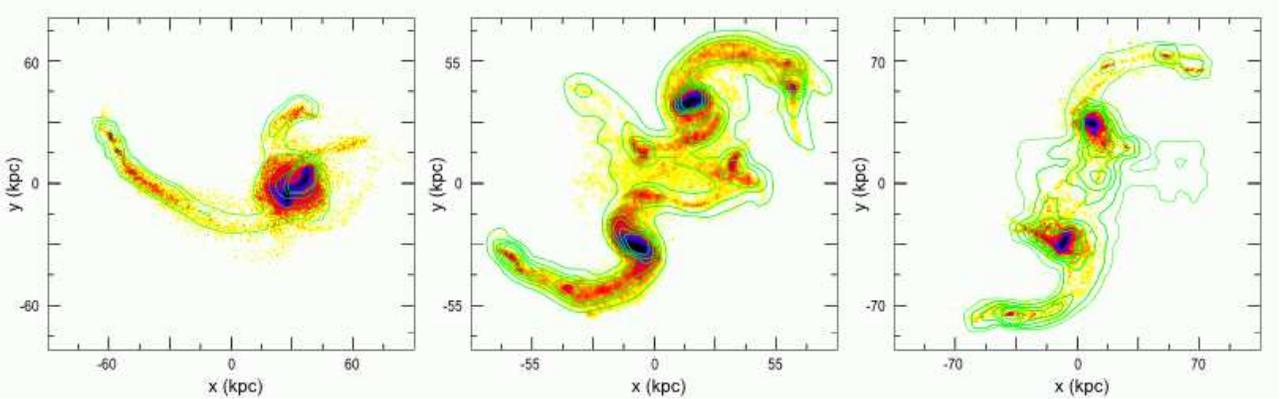,angle=-90,width=17cm}}
\caption{Results of the full N--body simulations of runs B (left), C (middle) and D
 (right) 50 to 100 Myrs after the periapse (see
Table~\ref{tab:TDGs}). Gas contours are superimposed on the stellar component.
All systems are seen face-on.}
\label{fig:simBCD} 
\end{figure*}

Fig.~\ref{fig:simBCD} presents snapshots 50 to 100 Myr after the periapse for runs B, C and D which were all
carried out with the extended dark matter haloes. 
Despite their slightly different initial conditions, orbital
parameters and a spatial resolution which is twice lower, they all show at the tip of the tails
 prominent gaseous condensations 
with masses well above $1.0 \x 10^9~\Mo$. Table~\ref{tab:TDGs} lists their properties.
In run B, the longest tail gives birth to a gravitationally bound TDG, while the shortest tail exhibits a smooth accumulation of matter at its tip which is not gravitationally bound and only transient. 
In run C, one of the major tidal accumulations fragmented into two pieces; the other one remains in the form of a single 
object. The two countertails (near $y=0$ on Fig.~\ref{fig:simBCD}) also show accumulations of matter in their outer parts, 
but these ones do not survive as they fall back quickly onto the parent galaxies. Run D, as well as run C, shows on top of the prominent TDG progenitor a few small stellar condensations, distributed along the tail, with masses of the order of 
$10^8~\Mo$ for the most massive ones; 
these gravitational clumps have similar masses as those  produced in the simulations by \cite{Barnes92} and \cite{Elmegreen93}. Like in  \cite{Elmegreen93},  they were
originally formed in gaseous condensations and  generally less  than 20-- 25\% of their mass correspond to  old stars  expelled from the parent spiral disk.
A few dynamical times later, they mostly consist of young stars formed in
situ from the gas clumps. Hence they  may be the
progenitors of  the SSCs observed in real interacting systems.
However, a detailed comparison of the simulated and real
 SSC is beyond the scope of this paper.

We carried out a few other simulations, truncating this time the
dark matter haloes. Sub-structures form along the tidal tails, but no
massive TDG progenitor is produced. Because these simulations are totally
similar to those carried out by the other groups \citep[e.g.,][]{Barnes92}, they are not presented 
here.


\begin{table}
\begin{minipage}{\columnwidth}
\centering
\begin{tabular}{llcccc}
\hline \hline
Run & Location & T \footnote{time after the periapse.} & $M_{\rm tot}$ \footnote{total mass.} & $M_{\rm *}^{\rm old}$ \footnote{mass of the old stellar population from the parent's disk.} & $M_{\rm *}^{\rm young}$ \footnote{mass of the young stellar population born in situ. Note that this value is approximate as the measurement of the SFR is 
affected by the limited numerical resolution.}\\
     & & (Myr) & (\Mo) & (\%) & (\%) \\
\hline
A & x$>$0 & 60 & $2.9 \x 10^9$ & 5 \% & 19 \% \\
   & x$<$0 &    & $2.3 \x 10^9$ & 12 \% & 13 \% \\
   & x$>$0 & 430 & $2.2 \x 10^9$ &  5\%  & 44\%  \\
   & x$<$0 &   & $1.8 \x 10^9$ & 14 \% &  31\% \\ \hline
B  & x$<$0 & 100 & $2.3 \x 10^9$ &  7\%  & 19\%  \\ \hline
C & x$>$0 & 90 & $2.5 \x 10^{9}$ \footnote{in two fragments.} &  11\%  & 16\%  \\
   & x$<$0 &  & $1.8 \x 10^9$ &  11\%  & 16\%  \\ \hline
D & y$>$0 & 55 & $1.1 \x 10^9$ &  3\%  & 14\%  \\
   & y$<$0 & & $2.4 \x 10^9$ &  8\%  & 19\%  \\ 
\hline
\end{tabular}
\end{minipage}
\caption{TDGs progenitors. }
\label{tab:TDGs}
\end{table}

Therefore, including the dynamical friction in full N-body simulations and increasing their resolution 
 did not change the basic results regarding the formation of TDGs initially presented in BDM03: 
massive gaseous clouds are still formed near the end of the tidal tail in this, otherwise, much
more realistic model.


\section{Mechanism for the formation of TDGs and role of the dark haloes}
\label{sect.:ingredients}

We have found that dark haloes should be very extended to enable the formation of massive TDGs.
Yet, since a more extended halo 
contains more mass, one may wonder whether the total DM mass is more important for the formation of the TDG precursors 
than its actual extent. We checked  that in  BDMD03 with simulations in which the truncated DM halo
 had a mass similar to the extended one. We were then unable to produce any massive accumulation in the tidal tails. Thus, 
the DM halo extent - and hence the shape of the corresponding potential well -- appears as the critical parameter, rather than the mass of the DM halo. 
Dubinski, Mihos \& Hernquist (1996)
\nocite{Dubinski96} and \cite{Mihos98b} actually claimed that massive, dense, haloes are less able to form
 long-lived tidal tails and hence proto--TDGs.  \cite{Mihos98b} noted however
that it is actually ``the shape and gradient of the galactic potential", rather than the mass of the DM halo, that 
determine the formation and  evolution of tidal tails.

In order to decipher the various mechanisms relevant for the formation of TDGs and weight their relative importance, we carried out a new set of 
restricted N-body simulations. The full N-body simulations  described in Sect.~\ref{sect.:hires-sim} 
ensured that, using  simpler models that can be more  easily analyzed, one would not disregard the main physical process.
In this whole section, we denote as {\it truncated} a DM halo truncated at three times the stellar disk radius, and 
{\it extended} a DM halo extended up to ten times the stellar disk radius. 

\begin{figure*}
\centerline{\psfig{file=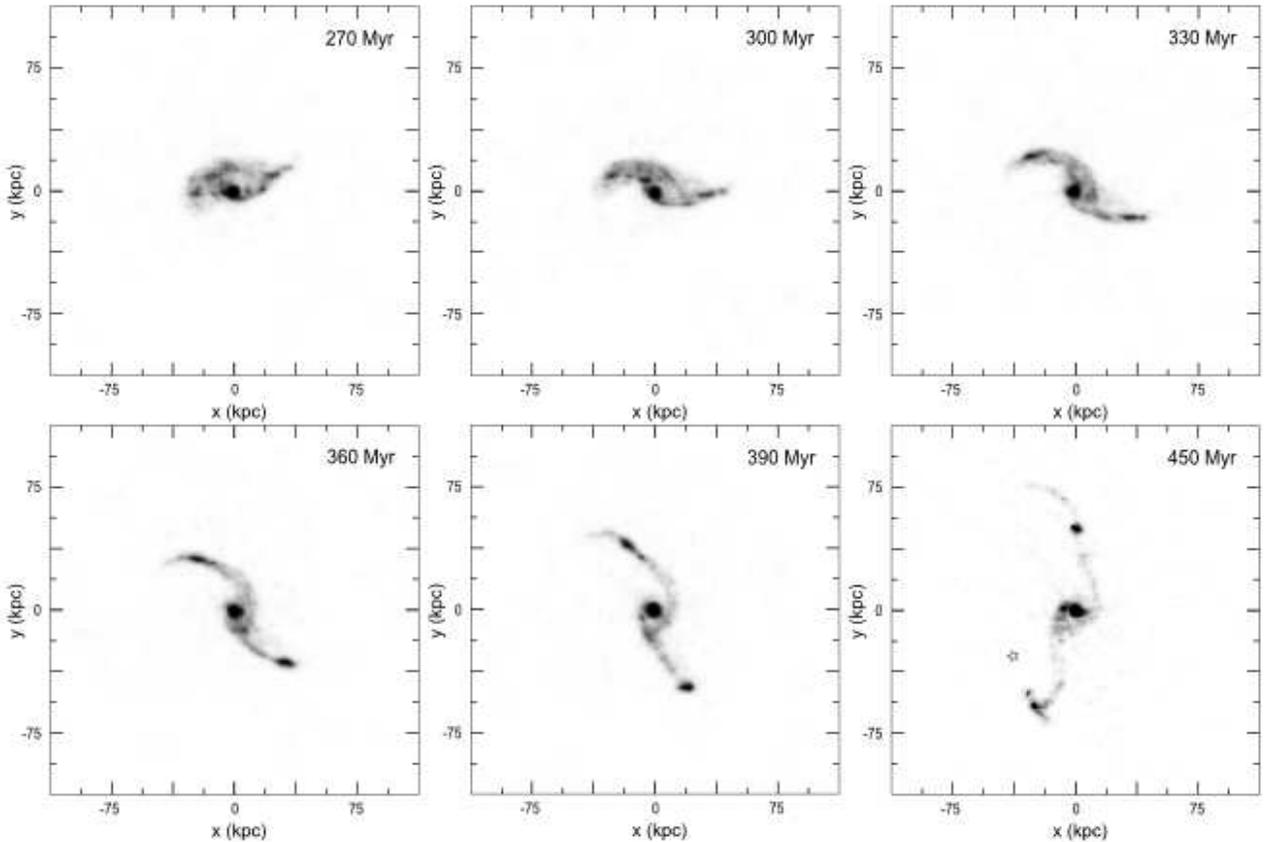,angle=-90,width=17cm}}
\caption{Formation of the TDG progenitors in the simulations carried out with the restricted N-body + rigid haloes code. 
The evolution of the gas distribution in one of the merging galaxies is presented. The position of the the second galaxy is shown by a star at $t=$450 Myr. At earlier
times, it is outside  the field of view. 
Self-gravity and energy dissipation in the ISM are included in these simulations.
The parameters of the simulation were: mass ratio 1:1, 8 \% of gas, impact parameter 150 kpc, initial relative velocity 125 km.s$^{-1}$.
}
\label{fig:3body-seq}
\end{figure*}

\begin{figure*}
\centerline{\psfig{file=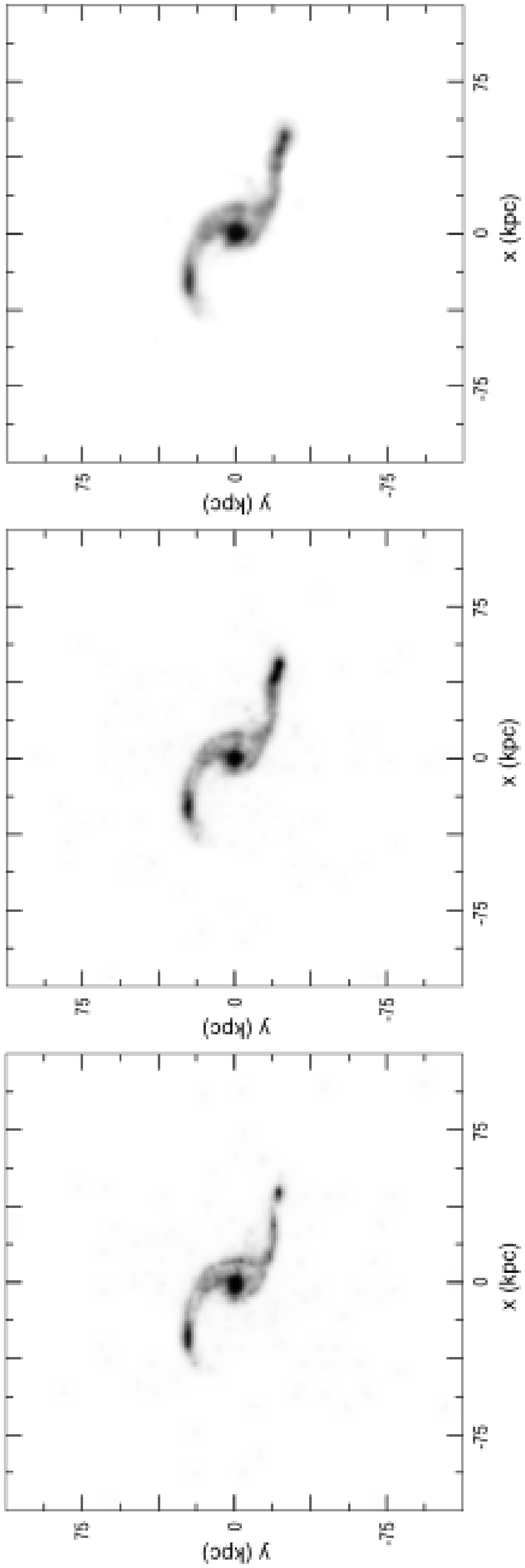,angle=-90,width=\textwidth}}
\caption{Left: TDG gaseous progenitors at $t=$345 Myr for the collision shown in Fig.~\ref{fig:3body-seq}. Middle: same simulation 
suppressing the self-gravity in the gaseous component; right: same simulation suppressing self-gravity and energy dissipation.}
\label{fig:3body-para}
\end{figure*}

\subsection{A kinematical origin}
\label{Sect.:kin}
We first analyze the role of dissipation and self-gravity using a hybrid code: N-body for the gas, and 3-body for the other
components.
In 3-body codes, massless  particles evolve in the rigid potential of two galaxies \citep{Toomre72}. Each galaxy potential 
was represented by a rigid rotation inside the central kpc, by a flat rotation curve for radii from 1 to 150 kpc, and a Keplerian 
potential at radii larger than 150 kpc. Such a profile corresponds to the case where the dark matter halo is extended.

The orbit of these two rigid systems were computed assuming that:
\begin{itemize}
\item the force exerted by a rigid system (G1) on the other galaxy (G2) can be determined at the mass center of G2.
\item the dynamical friction can be modeled by a viscous--like fluid drag  force proportional to the local density \citep[e.g.,][]{Combes95}.
 The force undergone by G1 is then assumed to be proportional to the density of G2 at the center of 
G1 $\rho_2(\vec{r_1})$, and to $\vec{v_2} - \vec{v_1}$, where $\vec{v_i}$ is the velocity of G$_i$. The force is then: 
$\vec{F}= k \rho_2(\vec{r_1})$ (\vec{v_2} - \vec{v_1}). The proportionality factor $k$ was chosen for the dynamical friction
 time-scale at distance 50 kpc to be 150 Myr, which is consistent with the results of previous N-body simulations. This 
description of the dynamical friction may seem very approximated, but N-body codes with a limited number of particles,
 and an arbitrary distribution of matter, also provide only an approximated friction.
\end{itemize}
We restricted our analysis to coplanar encounters, so that the gas particles evolve in a plane. In addition to the forces exerted on the two rigid galactic potential on gas particles in the frame of the  three-body method, we used:
\begin{itemize}
\item a bidimensional FFT code to account for the gas self-gravity
\item a sticky-particles algorithm to model the dissipative dynamics of the ISM
\end{itemize}
Both codes are described in \cite{Bournaud02}. We simulated galactic encounters with the same mass parameters for dark matter and the 
same initial gas distribution as in the full N-body simulations described in Sect.~2.

Fig.~\ref{fig:3body-seq} presents an example of a sequence of merging equal mass galaxies. Even in these simpler simulations, massive gaseous TDG progenitors of more than $10^9$ M$_{\sun}$ are formed. Note that, in this run, the condensations are not necessarily located at the very tip of the tidal tails. At time $t=$450 Myr, the condensation at $y>0$ is at about 2/3 of the tail extremity. This is also observed in real interacting systems, such as the prototype merger NGC 7252 \citep{Hibbard94}. Its HI tidal tail extends to the West further than its TDG  candidate, located at the tip of the optical tail. 
 
In Fig.~\ref{fig:3body-para}, we show the effect of switching off the self-gravity and the gas dissipation, and hence the ability of gaseous tidal condensations to  collapse. Large accumulations of gas are still present although they expectedly appear less compact and will probably not be able to survive for a long time in the numerical model. This clearly shows that the formation of massive TDGs is fundamentally a kinematical process and not the result of local gravitational instabilities that would later-on grow through accretion of surrounding material. The TDG progenitors are formed early on and already contain all the mass of the future TDG. Tidal forces within the potential well of extended dark matter haloes account for the large accumulations of tidally expelled gas. 
Only then, dissipation and self-gravity take over and trigger the cloud concentration and collapse.

One should note that, because of the kinematical origin of the proto--TDGs, our results are not sensitive to
 the fine-tuning of the gaseous dissipation rate which is, in numerical simulations,  often critical to structure
the tidal tails.


\subsection{The role of the DM halo extent}
\subsubsection{Response of a cold annulus to tidal forces}
In order to precisely understand the role of the DM
haloes around the perturbing/perturbed galaxy, we analyzed how an annulus of cold material reacts to
tidal forces. The  simulations were carried out with the same code 
and parameters as in Sect.~\ref{Sect.:kin}. Self-gravity has been disabled as we were only
interested in studying the kinematical aspects of the TDG formation.

\begin{figure}
\centering
\resizebox{7.2cm}{!}{\includegraphics{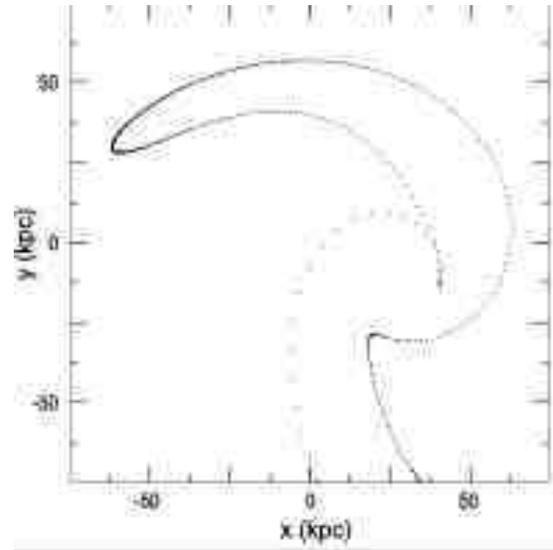}}
\caption{Response of an annulus of uniformly distributed , kinematically cold,  gas clouds to a tidal
perturbation.}
\label{fig:annulus}  
\end{figure}

\begin{figure}
\centering
\resizebox{9cm}{!}{\includegraphics{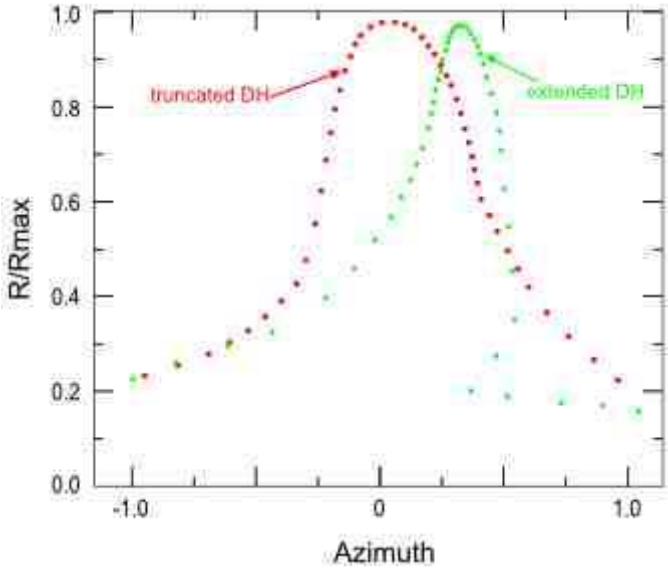}}
\caption{Response of an annulus of uniformly distributed gas clouds to a tidal
perturbation with  DM haloes as extended as ten stellar radii (green solid line) and with  
DM haloes truncated at three stellar radii (red dashed line). The x-axis corresponds to the 
azimuth after correction of the differential rotation, so that the tidal tail has a fixed azimuth equal to
zero. The radii $R$ have been scaled to the maximum extent of the tidal tail, $Rmax$.}
\label{fig:annulus-gaz}  
\end{figure}

We first considered an annulus made of gas with zero velocity dispersion located 
at a constant radius $R_0$ in the parent galaxy. It is perturbed by the tidal forces exerted by
the companion galaxy. We examined how particles that were initially uniformly distributed
along the annulus react to the perturbation 380 Myr after the beginning of the simulation, i.e. 
once the TDG progenitor is already visible, but before its collapse (see Fig.~\ref{fig:3body-seq}). 
As shown in Fig.~\ref{fig:annulus}, the annulus is transformed into a bent curve
 delineating the typical shape of a tidal tail.
To make the analysis easier, we subtracted the effect of the differential rotation in the parent galaxy.
In that new frame, tidal tails appear as linear 
structures, i.e. they have a mean constant azimuth, instead of being curved. In Fig.~\ref{fig:annulus-gaz},
we show the resulting plots in case of resp. extended and truncated DM haloes around the
parent and perturbing galaxies. 
 In both cases, the tail gets denser at its extremity, i.e. the particles look closer to each other
in the 1--D space examined here. 
The amplitude of the overdensity, however, depends on the size of the DM haloes
and is maximum when the haloes of both the parent galaxy and of the perturber are extended.
The compression corresponds to the decrease of the radial distance ($dr$) and more
significantly to the tightening of the azimuthal separation ($d\theta$) between the particles. 
In our simulations, $d\theta$ is three times smaller when the haloes are extended. 



\subsubsection{Amplitude of the radial excursions}

\begin{figure}
\centering
\resizebox{9cm}{!}{\includegraphics{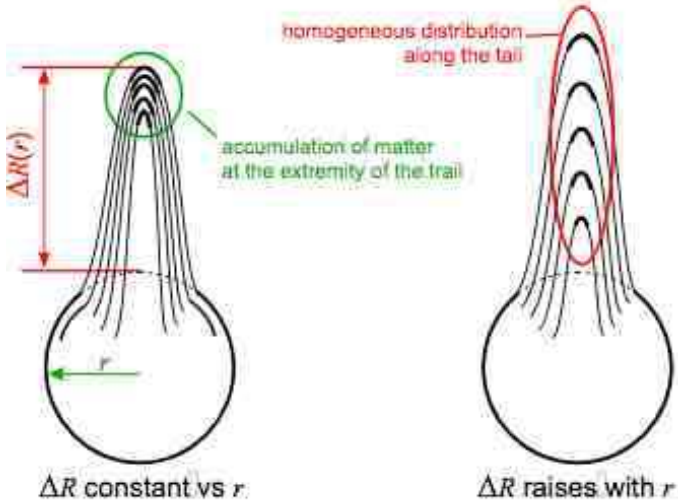}}
\caption{Schematic view of the effects of tidal perturbations to a series of concentric annuli 
of radius R for  extended (left) and truncated (right) dark matter haloes. The annuli were initially
 regularly spaced out.}
\label{fig:deltaR:sketch} 
\end{figure}

\begin{figure}
\centering
\resizebox{9cm}{!}{\includegraphics{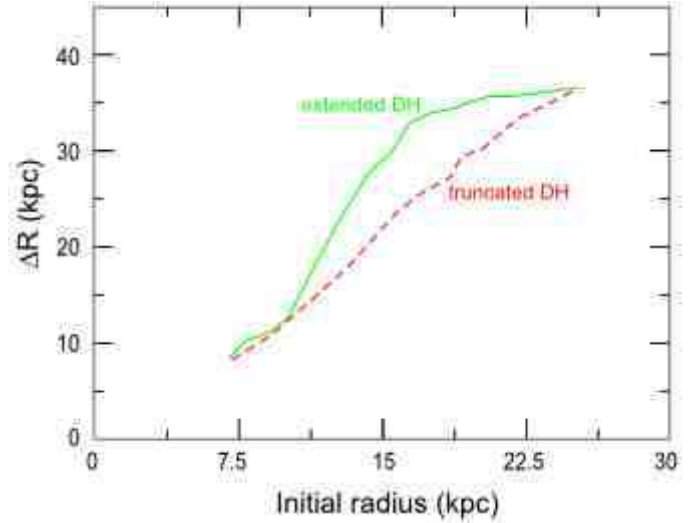}}
\caption{Amplitude of the radial excursions as a function of the initial radius with  DM haloes as 
extended as ten stellar radii (green solid line) and with  DM haloes truncated at three stellar radii (red dashed line).}
\label{fig:deltaR}  
\end{figure}

We then considered the more realistic situation of a set of concentric annuli with
initial radii $R_0$ perturbed by tidal forces. We examined how the amplitude of
the radial excursion, i.e. the difference $\Delta R$ between the maximum radial 
extent (reached at the azimuth of the tidal tail) and the initial radius 
(see Fig.~\ref{fig:deltaR:sketch}), varies with $R_0$. 
This will tell whether the 1--D overdensities observed for each single annulus overlap and 
hence form a 2--D condensation or whether they are uniformly distributed along the tail, canceling out 
any density contrast.
Fig.~\ref{fig:deltaR:sketch} illustrates the two situations.
In Fig.~\ref{fig:deltaR} we show the evolution of $\Delta R$ as a function of $R_0$
for the truncated and extended haloes.
In the first case, $\Delta R$ is a linear function of $R_0$, whatever $R_0$. 
Thus, the separation between two annuli in the tail is amplified and the overdensities 
are stretched along the tails.
On the other hand, for extended haloes, $\Delta R$ is nearly constant beyond a 
certain initial radius (15 kpc in the simulation shown in Fig.~\ref{fig:deltaR}). Thus the distance between the
annuli in the external regions of the parent galaxy is conserved
 and the overdensities are not diluted. 
Below this radius, the slope of the curve $\Delta R(R_0)$ is higher
than in the truncated case, and hence, the matter is even more stretched along
the tail. Both effects contribute to enhance the density contrast at the tip of the
tail, and account for the accumulation of large quantities of matter there.
In our simulations, the stretching along the radial direction, $ d \Delta R (Ro) / d Ro $, 
 is reduced by a factor of at least 7 (i.e.,
the compression enhanced by this value) for the 5 external kpc of the initial
disk when the haloes are extended.

\vspace{0.5cm}
In summary,  extended dark haloes cause, for a given annulus,  a compression of matter at the extremity of the
future tidal tail (smaller $d \theta$), and a smaller separation between  several concentric annuli 
(smaller $ d \Delta R (Ro) / d Ro $). Thus,  matter appears more concentrated both in the azimuthal and radial 
directions  than with truncated dark haloes. 

The physical explanation for the kinematical formation of a TDG progenitor
 lies in the shape of the tidal forces exerted on matter depending on its location with respect to the DM haloes. 
A mathematical description of the tidal field and its effect on test particles
is given below.

\subsection{The shape of the tidal field}
\label{sect:TF}

The potential field outside of a truncated halo is a Keplerian
potential, as arising from a point-like mass, while the potential
field inside of an extended halo can be accounted for, with a good
approximation, by an isothermal sphere potential. One reason why a
Keplerian potential and an isothermal sphere potential lead to so
qualitatively different outcomes is likely linked to the properties
of the tidal field of a spherical potential $\phi(r)$ \citep*{Dekel03}.

Assuming the potential center lies at $X=-r_0$, $Y=0$ and $Z=0$ in a
cartesian $(X,Y,Z)$ frame, we can evaluate the corresponding tidal field defined as
the differential acceleration between a test particle located at an
arbitrary position $X,Y,Z$ and a test particle located at the frame
origin. To lowest order in  $X/r_0$, $Y/r_0$ and $Z/r_0$,  the
components of the tidal field read:
\begin{eqnarray}
\label{eqn:tide}
\gamma_X&=&-(\partial_{r^2}^2\phi) X\nonumber\\
\gamma_Y&=&-(1/r_0)(\partial_r\phi) Y\nonumber\\
\gamma_Z&=&-(1/r_0)(\partial_r\phi) Z.
\end{eqnarray}
From these equalities one can see directly that for a Keplerian
potential ($\phi\propto r^{-1}$) one has
$\gamma_Y/Y=\gamma_Z/Z=-(1/2)\gamma_X/X$, while for an isothermal
sphere potential (for which $\phi\propto \log r$), one has:
$\gamma_Y/Y=\gamma_Z/Z=-\gamma_X/X$. In particular, if one is
interested in motions restricted to the $(X,Y)$ plane, then the tidal
field is divergence free for an isothermal sphere potential (i.\
e. inside a dark matter halo), while this divergence is positive for a
Keplerian potential (i.\ e. outside of a dark matter halo). In other
words, the tidal field of a Keplerian potential tends not only to
stretch fluid elements, but also to dilute them, whereas the tidal field
of an isothermal sphere tends to stretch fluid elements while
conserving their surface density.

 In order to further investigate this role
of the tidal field shape, we have performed simple calculations of a
cold bidimensional disk of test particles orbiting within its own
extended halo represented by an isothermal sphere potential (i.\
e. with a flat rotation curve). It is perturbed by an external, time varying,
tidal field, as would arise from a perturbing galaxy. 

The tidal field is expressed by:
\begin{eqnarray}
\label{eqn:tidalfield}
\gamma_X&=& \lambda_1 f(r) X\nonumber\\
\gamma_Y&=& \lambda_2 f(r) Y,
\end{eqnarray}
where the frame $(X,Y)$ is centered on the disk and has its $X$-axis
oriented toward the fictitious by-passing perturber located at distance
$r(t)$ from the center of the disk. The two numerical constants
$\lambda_1$ and $\lambda_2$ describe the shape of the tidal field and
therefore correspond to its local properties, while the
function $f(r)$ describes how the tidal field intensity decays with
the distance to the perturber ($f(r)\propto r^{-2}$ for an isothermal
sphere potential, while $f(r)\propto r^{-3}$ for a Keplerian
potential), and is therefore a global property of the tidal field.
The $\lambda_1$ and $\lambda_2$ constants are actually imposed by the
$f(r)$ function, through the Eqs~(\ref{eqn:tide}). It is however 
instructive to artificially set $\lambda$ and $f(r)$ separately in order to 
disentangle the different effects at work in exciting the tidal arms. 

\begin{figure}
\centering
\resizebox{\columnwidth}{!}{\includegraphics{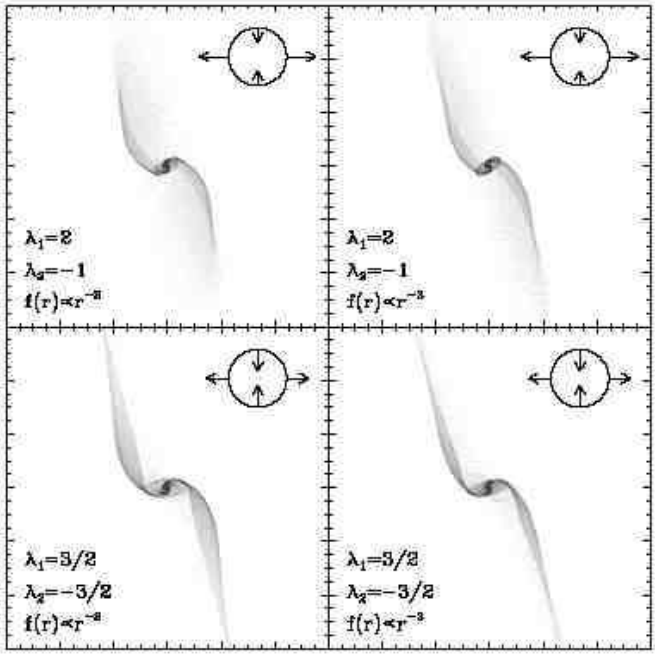}}
\caption{\label{fig:toymodel}
Plots in the left column correspond to the decay of an extended halo
potential ($f(r)\propto r^{-2}$) while those of the right column
correspond to a Keplerian decay ($f(r)\propto r^{-3}$) of the tidal
field. Similarly, the runs of the top row correspond to a local shape
of the tidal field that is Keplerian ($\lambda_1=-2\lambda_2$), while
the runs of the bottom row correspond to a local shape of the tidal
field corresponding to an isothermal sphere potential
($\lambda_1=-\lambda_2$). By comparing these four plots, one can
conclude that it is the relative value of $\lambda_1$ and $\lambda_2$
that primarily determines the arms morphology, i.\ e. the local shape
of the tidal field. The grey levels correspond to the same values on
the four plots, and scale with the square root of the surface
density. Notice in particular the presence of a diffuse scattered
population of test particles in the cases where the tidal potential has
a Keplerian shape (top row). The symbols in each panel represent the 
shape of the tidal field as in \cite{Dekel03}}
\end{figure}

Namely, we have run several calculations for
which we either set $\lambda_1=2,\lambda_2=-1$ (which corresponds to a
Keplerian tidal field shape), or $\lambda_1=3/2,\lambda_2=-3/2$ (which
corresponds to an isothermal sphere tide shape), everything else being
fixed (in particular the $f(r)$ function). We find with this toy model that the tidal arms
corresponding to this second choice are systematically more pronounced
and longer than the tidal arms excited by a Keplerian tidal field.
 In a similar manner, we have performed calculations in which we either
have $f(r)\propto r^{-2}$ or $f(r)\propto r^{-3}$, everything else
being fixed. Although the tidal arms position angle might vary
slightly between two such calculations, they
display  similar lengths and more significantly similar surface densities, which enables us
to conclude that it is primarily the shape of the tidal field (i.\ e.
the relative value of $\lambda_1$ and $\lambda_2$) that is
important in shaping and structuring the tidal arms. Fig.~\ref{fig:toymodel}
illustrates this conclusion. It represents at the same date the
surface density of a disk perturbed by a tidal field of the type given
by Eq.~(\ref{eqn:tidalfield}), with different prescriptions for 
$(\lambda_1,\lambda_2)$ and $f(r)$.\\

\cite{Wallin90} made a similar study of the structure of tidal tails, based on restricted 3-body 
simulations. He noted a density increase in
the tidal tails caused by their twisting and the formation of orbital caustics on their
outside edge. Such caustics which correspond to the crossing of particle
trajectories are also visible in Fig.~\ref{fig:toymodel}. Like a density wave, the
twist creates a local overdensity which propagates outwards. 
Meanwhile the amplitude of the density enhancement decreases. 
\cite{Wallin90} explored a wide variety of initial conditions and orbital parameters,
 but did not explicitly examined how his results were affected by the shape of the tidal 
field, as influenced by the size of the DM haloes.
 He speculated that, when the mass distribution is extended, the orbital
speed of the companion decreases, and the
propagation of the twist is delayed and slowed down.

Our simple calculations however strongly suggest that it is the
special shape of an extended halo tidal field ($\lambda_1=-\lambda_2$) that is
crucial to enable the formation of TDGs.

These results agree with what we previously found: when the tidal interaction occurs outside 
truncated dark haloes, the divergent tidal field inhibits the formation of massive accumulations of matter. 
When the interaction occurs inside extended dark haloes, the tidal field can efficiently carry away 
from the disk a large fraction of the gas, while maintaining its surface density to a high value. 
This results
in long, contrasted tidal tails, at the end of which further processes
can lead to the formation of TDG--like objects.


\section{Discussion}
\label{sect.:discussion}

\subsection{Top-down versus bottom-up scenario}
\label{sect:topdown}
The formation mechanism of classical galaxies has long been a puzzle. Whether they form from
the merging of smaller sub-structures, i.e. dwarf galaxies, -- the bottom-up scenario --,
or from the collapse of larger structures -- the top--down scenario --, has been actively debated.
The widely supported Cold Dark Matter model now promotes the hierarchical scenario where
small objects form first, although a monolithic collapse is still regularly claimed, especially to account 
for the formation of the old ellipticals in clusters. 

Even if Tidal Dwarf Galaxies have a very different origin, the discussion
on their formation mechanism may, to a large extent, be brought up in similar terms.
The early numerical simulations by \cite{Barnes92} and \cite{Elmegreen93} 
showed the formation of bound structures
all along the  tidal tails, with typical masses of $10^{7} - 10^{8}$~\Mo. One may have 
wondered whether they were the progenitors of the more massive (and gaseous) TDGs
observed in some interacting systems. Indeed they could in principle accrete surrounding
tidal material, in particular gas clouds. However,  the uniform distribution  of the bound clumps along the tails argue against this scenario, at least for the massive TDGs
 found near the tip of the tails.

Our simulations indicate that, whenever the dark matter haloes around the parent galaxies 
are sufficiently extended (their radii are at least 10 times that of the optical radius, or 150
kpc for a typical spiral galaxy), disk material coming from outside a given radius
may pile up at or near the end of the tidal tails, instead of being stretched along them. 
Thus, a kinematical process, rather than a local growing instability, seems to be at the very origin
of the progenitors of massive TDGs. Because the gas is more extended than the stars in the 
parent's disk, the proto--TDGs are initially mainly made of gas. Later on, the clouds
collapse under the effect of gravity and start forming stars. At the end of our N-body simulations,
the bulk of the stellar population is made of stars born in situ in the tail. In a few runs, the progenitor
broke into two fragments with individual masses still above or around $10^{9}$~\Mo.
Therefore, the formation of TDGs is more consistent with a top-down scenario.


\subsection{Redefining Tidal Dwarf Galaxies}
Massive TDGs are not the only objects formed out of tidal debris. We enumerated in the 
introduction many other by-products of galaxy collisions, such as Super Star Clusters,
Young Globular Clusters, Giant HII Complexes, etc ... Many of them are actually
called TDG candidates in the literature. Overall, the definition of a TDG remains elusive.
A basic consensual characteristic of any object defined as a TDG is that it should be
a gravitationally bound system \citep{Duc00}, although such a property is, observationally,
difficult to assess \citep{Hibbard01}. High resolution kinematical studies may help to
find evidences for a kinematical independence of tidal objects 
\citep{Duc98b,Mendes01,Weilbacher03,Bournaud04}.
Now, not all self-gravitating structures in tidal tails have the required initial mass to form
a genuine galaxy. Taking into account evaporation and fragmentation processes plus
tidal disruption, a total mass as high as $10^{9}$~\Mo\ may be necessary. This is the typical 
mass of the giant HI 
accumulations observed near the tip of several long tidal tails. Less massive
condensations may evolve, if they survive, into objects more similar to globular clusters.

However, if all tidal objects were formed by the same physical mechanism, fixing a
mass threshold for defining a TDG would appear arbitrary. Beside, a weakness of such a definition is
that we do not a priori know how a proto--TDG evolves. Numerical simulations
have just now the required resolution to study the precise fate of tidal debris (see next 
section), while, observationally, no 'old', independent TDG, detached from the parent 
galaxies, has yet been unambiguously identified \citep{Duc04a}. Their final mass and morphology 
-- a dwarf spheroidal, as suggested by \cite{Kroupa98} ? -- is then questionable.

The results of our simulations and their  modeling provide a stronger theoretical
 support for distinguishing the massive gas accumulations near the tip of tidal
tails and the other less massive condensations along them. Indeed, as stated in
Sect.~\ref{sect:topdown}, they seem to have formed by two different mechanisms.
We therefore wish to define as proto--TDGs  only the objects of the first 
category, i.e. those made of tidal material that was able to accumulate in the
tail and later collapse, according to a top-down scenario. 
Indeed, this process is the only one able to form condensations with masses greater
than $10^{9}$~\Mo, that are enough distant ($>$ 50 kpc) from their parent galaxies to avoid
a rapid fall back \citep{Hibbard95b} or a tidal disruption. 
In contrast, the clumps along the tails have basically a gravitational origin 
and formed from instabilities in either the stellar \citep{Barnes92} or
gaseous \citep{Elmegreen93} components.

Adopting this definition, one should admit that only specific, restrictive, conditions are 
required to form a TDG, in particular orbital parameters of the collision that favor
the development of long tidal tails, and the presence of extended dark matter haloes
around the parent galaxies. Both criteria may actually be simultaneously fulfilled.
Indeed \cite{Springel99} noted that less extensive haloes make weaker tails.
Another consequence is that only one proto--TDG 
per tail may be formed. However, it may later fragment (as seen in Fig.~\ref{fig:simBCD})
and generate a few extra TDGs. Determining the production rate of TDGs in merging
systems is beyond the scope of this paper. It would require to systematically
explore the parameter space of the collisions (varying the orbital 
parameters, structural properties of the parent galaxies, their mass ratios, etc ...).
We are currently engaged in such a long term program.

Do these conditions exclude that TDGs and the less massive SSCs form 
simultaneously ? This is not necessarily the case. In the simulations shown
in Fig.~\ref{fig:simBCD} (especially run D), a few faint  clumps can be seen along one
 of the the tails. 
They look like the bound objects produced in the simulations by  \cite{Elmegreen93}, although they are much less numerous in our simulations.
 Indeed, because a TDG already collects
a large fraction of the tidal material (about 75\%),
 there remains much less of it for the SSCs progenitors. 
This would be consistent with the observation by \cite{Knierman03} that
massive and compact condensations do not cohabit within a single tidal tail.

\subsection{Evolution of the proto--TDGs}
In the preliminary low resolution simulations presented in BDM03, we noted that the 
TDGs survive for more than 1.5 Gyr and end up orbiting around the merger remnant,
like any satellite galaxy. This result is in agreement with the study of \cite{Hibbard95b}
who, studying the return of tidal material in the prototype merger NGC 7252, concluded
that particles initially sent away at more than 100 kpc would not fall back within a
Hubble time.
 Our new full N-body simulations have a higher resolution, include 
feedback but were ended earlier. Still, we found that 0.5 Gyr after the formation of their progenitors, 
the two TDGs already appear isolated, orbiting at radial distances of about 100 kpc
(See fig.~\ref{fig:simA:face-on}). Meanwhile, they have already used up several tens of percent of their gas
reservoir and transformed them into young stars.
This stellar population born in situ dominates the older component tidally stripped
from the parent galaxies. Which confidence, however, can we give to these results on
the evolution of TDGs?

The resolution of our full N-body simulations is as a matter of fact still insufficient to firmly predict the 
 survival time of the TDGs. 
Each TDG progenitor is modeled by typically $10^{5}$ gas particles and
$10^{4}$ star particles. This is just enough to model the destructive feedback effect 
(due to its internal starburst). A refined grid around the TDG would be required to properly take into 
account the tidal shear in the young galaxy (induced by the parent galaxy) and, possibly, the fragmentation 
of its progenitor. Increasing the resolution of our simulations towards forming TDGs, we should hopefully soon 
infer the real impact of these additional processes on their evolution and life expectancy.

\section{Conclusions}

In this paper, we presented a set of numerical simulations of various types 
aimed at studying the
structure of tidal tails and their ability to concentrate prominent condensations likely to
become Tidal Dwarf Galaxies. Our immediate objective was to check the results
obtained by \cite{Bournaud03} with a simple numerical model: massive TDG progenitors 
may form provided that
the dark matter haloes around their parent galaxies are very extended, at least ten times 
more than the stellar disk.
Carrying out further tests, we reached the following conclusions:

\begin{itemize}
\item Gas accumulations as massive as $10^9~\Mo$ are still formed near the end of tidal tails 
 in N-body codes that include a large number of test particles, self-gravity, dissipation and
feedback. Our initial results were therefore robust. 
\item Switching off the self-gravity and dissipation in simpler  N-body simulations with rigid haloes
does not inhibit the gathering of tidal debris. This implies that the formation of massive tidal 
accumulations is fundamentally due to a kinematical process: tidal forces 
make a significant fraction of the gas coming from the outer parent's disk to pile up near the 
extremity of the tail. Once enough material 
has accumulated, self-gravity may lead to the local collapse of the gas clouds.
\item Using toy models, we clarified the role of the dark matter haloes around
the parent galaxies. We first analyzed how a 1--D annulus of cold gas that was originally uniformly 
distributed responds to a tidal perturbation. We observed an overdensity at the tip of the 
 tidal tail that is much more pronounced when the dark
matter haloes around both the perturbed galaxy and the perturber are extended.
We further determined the amplitude of the radial excursions exerted on a set of concentric
 annuli. For truncated DM haloes, the tidal material, and hence the individual 
1--D overdensities, are stretched all along the tail. On the other hand, when the DM haloes
are extended, this dilution only occurs below an initial radius. The density contrast
at the tip of the tail is then enhanced. 
\item Test particles expelled outside a truncated halo are subject to a Keplerian potential
while the particles that will remain inside the extended halo feel an isothermal 
sphere potential. The differences in the resulting tidal fields account for the
different efficiencies with which large quantities of gas may be carried away. 
We found that the local shape of the tidal field plays the most important role.
 \item The objects born in the prominent gas accumulations 
 were formed according to a top-down scenario, contrary to the less massive  
Giant HII complexes and Super Star  (Globular) Clusters that also  formed along tidal tails
 from growing local instabilities.
Having different physical origins, these different classes of collisional by-products have also
 different fates.
 \item We wish to define as Tidal Dwarf Galaxies (TDGs) only those objects having a 
progenitor exceeding $10^{9}~\Mo$, formed out of collapsing gaseous material that
accumulated far enough from the parent galaxies, usually (but not necessarily) 
near the tip of the tidal tails. Indeed, they are the most likely to survive and become genuine 
satellite galaxies. Such TDGs can only form under specific conditions, in particular
collisions that favor the development of long tidal tails within extended dark
matter haloes. They were observed so far in a limited number of interacting systems,
but a systematic census of such objects is still to be done.
\end{itemize}

More extensive simulations of galaxy collisions are required to assess the cosmological
 importance of TDGs. The parameter space of the orbits and initial physical conditions in the
parent galaxies should
be more thoroughly explored to estimate the production rate of the proto--TDGs. Using a more refine grid would help to study the evolution of the
TDGs and determine their life expectancy.

\begin{acknowledgements}
The numerical simulations were carried out on the NEC-SX6 vectorial computer of the CCRT 
of CEA.
We are most grateful to Romain Teyssier for stimulating 
discussions on these simulations and  to Fran\c{c}oise Combes for her advice on this project and
her comments on the manuscript. 
\end{acknowledgements}

\bibliographystyle{bib}
\bibliography{all}

\begin{thebibliography}{52}
\expandafter\ifx\csname natexlab\endcsname\relax\def\natexlab#1{#1}\fi
\expandafter\ifx\csname url\endcsname\relax
  \def\url#1{\texttt{#1}}\fi
\expandafter\ifx\csname urlprefix\endcsname\relax\def\urlprefix{URL }\fi

\bibitem[{{Barnes}(1992)}]{Barnes92c}
{Barnes}, J.~E., 1992, \apj 393, 484

\bibitem[{Barnes \& Hernquist(1992)}]{Barnes92}
Barnes, J.~E., Hernquist, L., 1992, Nature 360, 715

\bibitem[{{Barnes} \& {Hernquist}(1996)}]{Barnes96}
{Barnes}, J.~E., {Hernquist}, L., 1996, \apj 471, 115

\bibitem[{{Bekki} et~al.(2002){Bekki}, {Forbes}, {Beasley}, \&
  {Couch}}]{Bekki02}
{Bekki}, K., {Forbes}, D.~A., {Beasley}, M.~A., {Couch}, W.~J., 2002, \mnras
  335, 1176

\bibitem[{{Bournaud} \& {Combes}(2002)}]{Bournaud02}
{Bournaud}, F., {Combes}, F., 2002, \aap 392, 83

\bibitem[{{Bournaud} \& {Combes}(2003)}]{Bournaud03a}
{Bournaud}, F., {Combes}, F., 2003, \aap 401, 817

\bibitem[{{Bournaud} et~al.(2004){Bournaud}, {Duc}, {Amram},
  et~al.}]{Bournaud04}
{Bournaud}, F., {Duc}, P.~., {Amram}, P., et~al., 2004, \aap in press
  (astro-ph/0406169)

\bibitem[{{Bournaud} et~al.(2003){Bournaud}, {Duc}, \& {Masset}}]{Bournaud03}
{Bournaud}, F., {Duc}, P.-A., {Masset}, F., 2003, \aap 411, L469

\bibitem[{{Braine} et~al.(2001){Braine}, {Duc}, {Lisenfeld}, et~al.}]{Braine01}
{Braine}, J., {Duc}, P.-A., {Lisenfeld}, U., et~al., 2001, \aap 378, 51

\bibitem[{{Combes} et~al.(1995){Combes}, {Boisse}, {Mazure}, et~al.}]{Combes95}
{Combes}, F., {Boisse}, P., {Mazure}, A., et~al., 1995, {Galaxies and
  Cosmology}, Springer-Verlag Berlin Heidelberg New York.~ Also Astronomy and
  Astrophysics Library

\bibitem[{{Cortese} et~al.(2004){Cortese}, {Gavazzi}, {Boselli}, \&
  {Iglesias-Paramo}}]{Cortese04}
{Cortese}, L., {Gavazzi}, G., {Boselli}, A., {Iglesias-Paramo}, J., 2004, \aap
  416, 119

\bibitem[{{de Grijs} et~al.(2003){de Grijs}, {Lee}, {Clemencia Mora Herrera},
  et~al.}]{deGrijs03}
{de Grijs}, R., {Lee}, J.~T., {Clemencia Mora Herrera}, M., et~al., 2003, New
  Astronomy 8, 155

\bibitem[{{Dekel} et~al.(2003){Dekel}, {Devor}, \& {Hetzroni}}]{Dekel03}
{Dekel}, A., {Devor}, J., {Hetzroni}, G., 2003, \mnras 341, 326

\bibitem[{{Dubinski} et~al.(1996){Dubinski}, {Mihos}, \&
  {Hernquist}}]{Dubinski96}
{Dubinski}, J., {Mihos}, J.~C., {Hernquist}, L., 1996, \apj 462, 576

\bibitem[{{Dubinski} et~al.(1999){Dubinski}, {Mihos}, \&
  {Hernquist}}]{Dubinski99}
{Dubinski}, J., {Mihos}, J.~C., {Hernquist}, L., 1999, \apj 526, 607

\bibitem[{{Duc} et~al.(2004){Duc}, {Bournaud}, \& {Masset}}]{Duc04a}
{Duc}, P.-A., {Bournaud}, F., {Masset}, F., 2004, in: IAU Symposium 217,
  Recycling intergalactic and interstellar matter, eds. Duc, P.~A., Braine, J.,
  Brinks, E., ASP,  550 (astro--ph/0402252)

\bibitem[{{Duc} et~al.(2000){Duc}, {Brinks}, {Springel}, et~al.}]{Duc00}
{Duc}, P.-A., {Brinks}, E., {Springel}, V., et~al., 2000, \aj 120, 1238

\bibitem[{Duc et~al.(1997)Duc, Brinks, Wink, \& Mirabel}]{Duc97b}
Duc, P.-A., Brinks, E., Wink, J.~E., Mirabel, I.~F., 1997, A\&A, 326, 537

\bibitem[{{Duc} \& {Mirabel}(1998)}]{Duc98b}
{Duc}, P.-A., {Mirabel}, I.~F., 1998, \aap 333, 813

\bibitem[{Duc \& Mirabel(1999)}]{Duc99b}
Duc, P.-A., Mirabel, I.~F., 1999, in: IAUS 186: Galaxy Interactions at Low and
  High Redshift, eds. Barnes, J.~E., Sanders, D.~B., Kluwer: Dordrecht, ~61

\bibitem[{Elmegreen et~al.(1993)Elmegreen, Kaufman, \& Thomasson}]{Elmegreen93}
Elmegreen, B.~G., Kaufman, M., Thomasson, M., 1993, ApJ 412, 90

\bibitem[{{English} et~al.(2003){English}, {Norris}, {Freeman}, \&
  {Booth}}]{English03}
{English}, J., {Norris}, R.~P., {Freeman}, K.~C., {Booth}, R.~S., 2003, \aj
  125, 1134

\bibitem[{{Gallagher} et~al.(2001){Gallagher}, {Charlton}, {Hunsberger},
  et~al.}]{Gallagher01}
{Gallagher}, S.~C., {Charlton}, J.~C., {Hunsberger}, S.~D., et~al., 2001, \aj
  122, 163

\bibitem[{{Gerhard} et~al.(2002){Gerhard}, {Arnaboldi}, {Freeman}, \&
  {Okamura}}]{Gerhard02}
{Gerhard}, O., {Arnaboldi}, M., {Freeman}, K.~C., {Okamura}, S., 2002, \apjl
  580, L121

\bibitem[{{Gilbert} et~al.(2000){Gilbert}, {Graham}, {McLean},
  et~al.}]{Gilbert00}
{Gilbert}, A.~M., {Graham}, J.~R., {McLean}, I.~S., et~al., 2000, \apjl 533,
  L57

\bibitem[{{Hibbard} \& {Barnes}(2004)}]{Hibbard04}
{Hibbard}, J., {Barnes}, J.~E., 2004, in: IAU Symposium 217, Recycling
  intergalactic and interstellar matter, eds. Duc, P.~A., Braine, J., Brinks,
  E., ASP,  510

\bibitem[{Hibbard et~al.(1994)Hibbard, Guhathakurta, van Gorkom, \&
  Schweizer}]{Hibbard94}
Hibbard, J.~E., Guhathakurta, P., van Gorkom, J.~H., Schweizer, F., 1994, AJ
  107, 67

\bibitem[{{Hibbard} \& {Mihos}(1995)}]{Hibbard95b}
{Hibbard}, J.~E., {Mihos}, J.~C., 1995, \aj 110, 140

\bibitem[{{Hibbard} et~al.(2001){Hibbard}, {van der Hulst}, {Barnes}, \&
  {Rich}}]{Hibbard01}
{Hibbard}, J.~E., {van der Hulst}, J.~M., {Barnes}, J.~E., {Rich}, R.~M., 2001,
  \aj 122, 2969

\bibitem[{{Hibbard} \& {van Gorkom}(1996)}]{Hibbard96}
{Hibbard}, J.~E., {van Gorkom}, J.~H., 1996, \aj 111, 655

\bibitem[{Holtzman et~al.(1992)Holtzman, Faber, Shaya, et~al.}]{Holtzman92}
Holtzman, J.~A., Faber, S.~M., Shaya, E.~J., et~al., 1992, AJ 103, 691

\bibitem[{{Jungwiert} et~al.(2001){Jungwiert}, {Combes}, \& {Palou{\v
  s}}}]{Jungwiert01}
{Jungwiert}, B., {Combes}, F., {Palou{\v s}}, J., 2001, \aap 376, 85

\bibitem[{{Knierman} et~al.(2003){Knierman}, {Gallagher}, {Charlton},
  et~al.}]{Knierman03}
{Knierman}, K.~A., {Gallagher}, S.~C., {Charlton}, J.~C., et~al., 2003, \aj
  126, 1227

\bibitem[{{Kroupa}(1998)}]{Kroupa98}
{Kroupa}, P., 1998, \mnras 300, 200

\bibitem[{{L{\' o}pez-S{\' a}nchez} et~al.(2004){L{\' o}pez-S{\' a}nchez},
  {Esteban}, \& {Rodr{\'{\i}}guez}}]{Lopez04}
{L{\' o}pez-S{\' a}nchez}, {\' A}.~R., {Esteban}, C., {Rodr{\'{\i}}guez}, M.,
  2004, \apjs 153, 243

\bibitem[{{Mendes de Oliveira} et~al.(2004){Mendes de Oliveira}, {Cypriano},
  {Sodr{\' e}}, \& {Balkowski}}]{Mendes04}
{Mendes de Oliveira}, C., {Cypriano}, E.~S., {Sodr{\' e}}, L., {Balkowski}, C.,
  2004, \apjl 605, L17

\bibitem[{{Mendes de Oliveira} et~al.(2001){Mendes de Oliveira}, {Plana},
  {Amram}, et~al.}]{Mendes01}
{Mendes de Oliveira}, C., {Plana}, H., {Amram}, P., et~al., 2001, \aj 121, 2524

\bibitem[{{Mihos}(2004)}]{Mihos04}
{Mihos}, C., 2004, in: IAU Symposium 217, Recycling intergalactic and
  interstellar matter, eds. Duc, P.~A., Braine, J., Brinks, E., ASP,  390
  (astro--ph/0401557)

\bibitem[{{Mihos} et~al.(1998){Mihos}, {Dubinski}, \& {Hernquist}}]{Mihos98b}
{Mihos}, J.~C., {Dubinski}, J., {Hernquist}, L., 1998, \apj 494, 183

\bibitem[{Mirabel et~al.(1992)Mirabel, Dottori, \& Lutz}]{Mirabel92}
Mirabel, I.~F., Dottori, H., Lutz, D., 1992, A\&A 256, L19

\bibitem[{{Roberts} \& {Haynes}(1994)}]{Roberts94}
{Roberts}, M.~S., {Haynes}, M.~P., 1994, \araa 32, 115

\bibitem[{{Ryan-Weber} et~al.(2004){Ryan-Weber}, {Meurer}, {Freeman},
  et~al.}]{Ryan-Weber04}
{Ryan-Weber}, E.~V., {Meurer}, G.~R., {Freeman}, K.~C., et~al., 2004, \aj 127,
  1431

\bibitem[{{Saviane} et~al.(2004){Saviane}, {Hibbard}, \& {Rich}}]{Saviane04}
{Saviane}, I., {Hibbard}, J.~E., {Rich}, R.~M., 2004, \aj 127, 660

\bibitem[{{Schmidt}(1959)}]{Schmidt59}
{Schmidt}, M., 1959, \apj 129, 243

\bibitem[{{Schweizer} et~al.(1996){Schweizer}, {Miller}, {Whitmore}, \&
  {Fall}}]{Schweizer96}
{Schweizer}, F., {Miller}, B.~W., {Whitmore}, B.~C., {Fall}, S.~M., 1996, \aj
  112, 1839

\bibitem[{{Springel} \& {White}(1999)}]{Springel99}
{Springel}, V., {White}, S. D.~M., 1999, \mnras 307, 162

\bibitem[{Toomre \& Toomre(1972)}]{Toomre72}
Toomre, A., Toomre, J., 1972, ApJ 178, 623

\bibitem[{{Tran} et~al.(2003){Tran}, {Sirianni}, {Ford}, et~al.}]{Tran03}
{Tran}, H.~D., {Sirianni}, M., {Ford}, H.~C., et~al., 2003, \apj 585, 750

\bibitem[{Wallin(1990)}]{Wallin90}
Wallin, J.~F., 1990, AJ 100, 1477

\bibitem[{{Weilbacher} et~al.(2003){Weilbacher}, {Duc}, \&
  {Fritze-v.~Alvensleben}}]{Weilbacher03}
{Weilbacher}, P.~M., {Duc}, P.-A., {Fritze-v.~Alvensleben}, U., 2003, \aap 397,
  545

\bibitem[{{Whitmore} et~al.(1999){Whitmore}, {Zhang}, {Leitherer},
  et~al.}]{Whitmore99}
{Whitmore}, B.~C., {Zhang}, Q., {Leitherer}, C., et~al., 1999, \aj 118, 1551

\bibitem[{{Zepf} et~al.(1999){Zepf}, {Ashman}, {English}, et~al.}]{Zepf99}
{Zepf}, S.~E., {Ashman}, K.~M., {English}, J., et~al., 1999, \aj 118, 752

\end{thebibliography}

\end{document}